# Point Contact Spin Spectroscopy of Ferromagnetic MnAs Epitaxial Films.

R. Panguluri[1], G. Tsoi[1], B. Nadgorny[1], S.H.Chun[2,3], N. Samarth[2], I.I. Mazin[4]

[1]Department of Physics and Astronomy, Wayne State University, Detroit MI 48201

[2]Department of Physics and Materials Research Institute, The Pennsylvania State University, University Park, PA 16802

[3] Department of Physics, Sejong University, Seoul 143-747, Korea

[4] Code 6390, Naval Research Laboratory, 4555 Overlook Ave. Washington DC 20375

## Abstract

We use point contact Andreev reflection spin spectroscopy to measure the transport spin polarization of MnAs epitaxial films grown on (001) GaAs. By analyzing both the temperature dependence of the contact resistance and the phonon spectra of lead acquired simultaneously with the spin polarization measurements, we demonstrate that all the point contacts are in the ballistic limit. A ballistic transport spin polarization of approximately 49% and 44% is obtained for the type A and type B orientations of MnAs, respectively. These measurements are consistent with our density functional calculations, and with recent observations of a large tunnel magnetoresistance in MnAs/AlAs/(Ga,Mn)As tunnel junctions.

Contemporary interest in both metallic [1] and semiconductor spintronics [2,3] has revived a long standing awareness of the importance of charge and spin coupling [4,5], with recent focus on fundamental understanding of the injection, transport and relaxation processes of spin in semiconductors and their heterostructures [6,7,8]. Unlike all-metal spintronics devices wherein efficient spin injection has been demonstrated some time ago [9], spin injection from ferromagnetic metals into semiconductors [10] proved to be a more daunting problem [11], due in part to the small boundary resistance at the metal/semiconductor interface, or the "conductivity mismatch" [12]. This hurdle can be overcome by spin injection via magnetic semiconductor contacts [6], fully spin polarized metal ("half-metal") contacts, or tunnel contacts [13,14], providing spin-selectivity at the interface [15]. In this context, it is particularly important to obtain experimental measurements of spin polarization in ferromagnets, such as MnAs and (Ga,Mn)As that can be naturally integrated with technologically important semiconductors such as GaAs. Another interest in this compound stems from the fact that, according to the recent band structure calculations [16], the hypothetical zinc-blend structure of MnAs, might be a half-metal [17], or very close to a half-metallic phase transition [16].

MnAs – a ferromagnetic metal – is a particularly promising material for semiconductor spintronics since it is a relatively high temperature ferromagnet ($T_C$ = 320 K) that can be integrated with GaAs-based heterostructures via molecular beam epitaxy (MBE) [18,19]. Furthermore, the observation of a large tunnel magnetoresistance (~30% at ~5 K) in MnAs/AlAs/(Ga,Mn)As junctions suggests MnAs as a viable candidate for tunnel spin injection into GaAs [13]. The epitaxial growth of MnAs on (001) GaAs is known to produce two possible crystalline orientations of the NiAs phase, selected by specific growth conditions [18,19]: in ''type A'' samples, the growth axis is along $[\bar{1}100]$ MnAs and the c-axis is in the (001) plane of the GaAs substrate, while in ''type B'' samples the growth axis is along $[\bar{1}101]$ MnAs planes and the c-axis is at an angle of ~ 23° to the plane of the substrate.

Here, we report the results of Point Contact Andreev Reflection (PCAR) measurements of the ballistic transport spin polarization in both type A and type B ferromagnetic epilayers of MnAs grown on (100) GaAs. The experimental measurements are consistent with our results of density functional calculations in the generalized gradient approximation (GGA).

Spin polarization can be defined in a number of different ways, depending on the experiment in question. The spin polarization for the density of states at the Fermi level, $P_0$ is naturally associated with the photoemission spectroscopy experiments. Our interest here focuses on the *transport* spin polarization that can be often written in the form

$$P_n = \frac{N_\uparrow(E_F)V_{F\uparrow}^n - N_\downarrow(E_F)V_{F\downarrow}^n}{N_\uparrow(E_F)V_{F\uparrow}^n + N_\downarrow(E_F)V_{F\downarrow}^n} \qquad (1),$$

where $N\uparrow(E_F)$ and $N\downarrow(E_F)$, $V_{F\uparrow}$ and $V_{F\downarrow}$ are the densities of states and the Fermi velocities for majority and minority spin sub-bands respectively, and where $P_1$ and $P_2$ refer to the ballistic and diffusive regimes, respectively.

The PCAR technique [20,21] based on Andreev reflection [22] measures the degree of the spin polarization of the current in a ferromagnet (F) interfaced with a superconductor (S). As de Jong and Beenakker pointed out [23] some electrons from the majority band of the ferromagnet cannot find a partner in the minority band to propagate inside the superconductor as a Cooper pair. For this spin polarized fraction of the current Andreev reflection is prohibited, strongly affecting the contact conductance below the superconducting gap. The current transfer across F-S interface can be described in detail [24], using a generalized Blonder-Tinkham-Klapwijk (BTK) model [25] for both ballistic (mean free path $L$ is larger than the contact size $d$) and diffusive ($L<d$) point contacts. Despite many approximations in the derivation of the generalized BTK formula [24] (e.g. a δ-function interface potential which may not be always adequate for real systems [26]) it often provides a good description of the experimental data, performing well beyond its limit of applicability. For instance, it is known to produce good fits of the experimental conductance spectra with the interface resistance $Z = 0$, although within the BTK model itself $Z$ can never be equal to exactly zero because of the Fermi velocity mismatch [26,27]. We caution that the correct interpretation of such spin polarization measurements also requires a clear identification of the transport regime (diffusive *versus* ballistic). We experimentally establish the transport regime by evaluating the temperature dependence of the point contact and the resistivity of the metal (MnAs). Additionally, we measure the *phonon spectra* of the point contact for Pb to evaluate the contact quality.

The Pb and Sn probes for this study were fabricated by fine mechanical polishing of the thin superconducting rods, similarly to the technique described in Ref. 20. The tip and the sample were balanced between two well-polished sapphire plates by a set of metal springs with low thermal expansion coefficient. A shaft connected to a differential type screw was used to change the position of the top plate by about 10 μm per revolution with respect to the bottom sapphire plate to which the sample was attached, allowing good control over the contact resistance with approximately order of magnitude better sensitivity compared to the previous work [20], [28]. After the contact is established, the shaft can be disconnected from the probe/sample stage. This eliminates most of the temperature gradients in the system and significantly increases the stability of the contact. Importantly, this design allowed us to measure the temperature dependence of the contact in a broad temperature range, which was used for independent determination of the mean free path and the contact size as described below. All the transport measurements were made with a conventional four-probe technique, with the differential conductance d*I*/d*V* obtained by standard ac lock-in detection at a frequency of ~2 kHz within the temperature range 1.2 – 4.2 K. Additionally, we have recorded the second derivative $\frac{d^2I}{dV^2}$ to obtain the point contact phonon spectra.

The MnAs epilayers used in this study have thicknesses in the range between 40nm and 200nm, and were grown in an EPI 930 MBE system using standard solid source effusion cells. The substrate temperature was monitored using a thermocouple situated behind the substrate mounting blocks. The sample growth was monitored using reflection high-energy electron diffraction (RHEED) at 12 keV. All samples in this study were grown on non-vicinal GaAs (001) substrates that were prepared using *in situ* thermal cleaning under As overpressure. A 100 nm thick GaAs buffer was first grown at 600 $^{o}$C. Upon cooling in an As overpressure to a substrate temperature of 250 $^{o}$C, RHEED indicated a c(4x4) reconstructed GaAs surfaces. Type A samples were produced by either directly depositing MnAs on this surface, or on an *annealed* thin layer of low temperature grown GaAs (LT-GaAs). In contrast, for type B samples, the MnAs layer was deposited on the unreconstructed surface of an unannealed LT-GaAs layer. In some of the thicker type A films we observe the corrugated structure, similar to the one recently reported in Ref. [29]; yet the spin transport properties of α-MnAs phase in these films were found to be the same as in MnAs samples with no corrugation. The magnetic properties of the MnAs films used in this study are comparable to that observed by other groups [18,29], with the coercivity along the easy $[11\bar{2}0]$ magnetization axis (~200Oe) approximately two orders of magnitude lower than the coercivity along the hard [0001] MnAs axis (~3T). The anisotropy constants, extracted from the hard axis magnetization curve, are $K_1$ = 610 kJ/m$^3$; $K_2$ = 250 kJ/m$^3$.

PCAR measurements were carried out on several MnAs samples, with approximately 10 different point contacts measured for each sample using both Sn and Pb superconducting tips. We note that the geometry of the epilayer only permits measurements along the growth axis: namely, along the $[\bar{1}100]$ direction for type A samples and along the $[\bar{1}101]$ direction for type B samples. Fig. 1 shows several consecutive conductance curves for different Sn contacts to type A MnAs sample. The dI/dV characteristics of this series of contacts are very similar, essentially independent of the contact resistance, thus yielding practically the same values of the spin polarization. The spin polarization is extracted from the d*I*/d*V* data in Fig.1 using a modified BTK theory [24]. Since, as it will be shown below, all the measured contacts are clearly in the ballistic regime, we use exclusively the ballistic theory to analyze the data. A typical fit of one of the curves from Fig. 1 (with the contact resistance of 37Ω) is shown in Fig. 2(a), while for comparison Fig. 2(b) shows an example of the fit for a superconducting Pb point contact with a type B MnAs. For both of the curves in Fig. 2, as well as for the great majority of all the other contacts used in this paper, the barrier strength $Z$ turned out to be rather small ($Z \sim 0.1 - 0.15$). Thus the question of a possible $Z$-dependence of the values of the spin polarization raised recently [30] does not appear relevant for this study.

The average spin polarization values obtained from our measurements for type A MnAs is $P \sim 49 \pm 2$ %. The average $P$ for type B MnAs is $\sim 44 \pm 4$%, and might indicate the presence of the spin polarization anisotropy between the in-plane and out-of plane spin currents, which is expected from our theoretical estimates [31]. However, this is difficult to conclusively confirm, since there are larger errors in the measurements of the type B MnAs compared to the type A samples, and since the difference in $P$ may also arise from different microscopic structure of type B samples. A single crystal of bulk MnAs is needed to give a definitive answer to this important question.

We have independently determined the transport regime, by estimating the elastic mean free path $L$ and the contact size $d$, using the approach proposed in Ref. 32. The contact size $d$ is evaluated from the temperature dependence of the contact resistance at zero bias $R_c(T)$ at sufficiently high temperatures, where $R_c(T)$ is proportional to the resistivity $\rho(T)$, and $\rho(T)$ is the temperature dependence of MnAs film [33] (see Fig. 3). Then $d = \frac{d\rho/dT}{dR_c/dT}$ and the mean free path $L = \frac{\langle \rho l \rangle}{d}\left[R(0) - \frac{16}{3\pi}\frac{\langle \rho L \rangle}{d^2}\right]^{-1}$. Our probe design allowed us to measure the temperature dependence of the contact in a fairly broad temperature range. The estimates for MnAs type A film and Sn tip with the typical contact resistance $R_c \sim 10\,\Omega$ was done in the temperature range 200-245K with $dR_c/dT = 0.22\,\Omega$/K and $d\rho/dT = 0.35 \cdot 10^{-6}\,\Omega$·cm/K, resulting in $d$ = 15 nm and $L$ = 330 nm. These estimates confirm that the contact is well inside the ballistic regime, $L$>>$d$, fully justifying the use of the ballistic case to analyze the Andreev reflection data.

To further check the integrity of our point contacts with MnAs we have also measured the phonon spectrum $d^2I/dV^2$ of lead in the superconducting state, which is proportional to the Eliashberg function $\alpha^2F(\varepsilon)$ (Fig 4). The observed spectral peaks, shifted to higher energies compared to the normal state, are clearly visible and very close to the known phonon peaks of Pb [34], as well as to the ones obtained previously by the Cornell group in the nanocontact geometry with Fe and Co [21]. As shown in Ref. 35, the impurities in the contact area can change the phonon energies, and also modify the peak amplitudes and broaden the spectra. The ability to see good phonon spectra is commonly associated with negligible bulk scattering within the contact area [21]. This additional control of the quality of the point contacts gives us yet another confirmation of the ballistic nature of the contacts and confidence that the PCAR technique applied to MnAs with Pb (Sn) points is a viable experimental geometry.

In order to further understand our experimental results we have performed band structure calculations of MnAs in the NiAs phase (hexagonal or α-MnAs), using the Linear Muffin-Tin Orbital method (Stuttgart LMTO-TB code) and the Generalized

Gradient Approximation (GGA) for the exchange-correlation potential [36]. The calculated band structure is in general agreement with the previous results [37]. The main feature of the band structure is that Mn majority d band is almost fully occupied, so that the bands crossing the Fermi level in this spin polarization originate from As p states, and have sizeable velocity. In the minority spin channel, although the Fermi level crosses the bottom of the Mn d bands, most of the Mn d bands are empty. Correspondingly, the Fermi velocity in this channel is smaller by approximately a factor of four (see Fig. 5). This is reminiscent of the 3d metals, like Ni, and, similarly, leads to a *negative* spin polarization in terms of the DOS at the Fermi level, but to a *positive* polarization of the current [38].

The calculations yield the spin polarization for density of states $P_0$ = -15%. For the ballistic transport in the hexagonal MnAs plane perpendicular to the c-axis $P_{1\perp}$ = 42%, and parallel to the c- axis ([0001] direction) $P_{1\parallel}$ = 15%. For diffusive (Ohmic) transport $P_{1\perp}$ = 80% and $P_{1\parallel}$ = 43%. Importantly, the latter result implies that 80% of the current in plane in the bulk of MnAs is transferred by the majority spins. The in-plane Fermi velocities calculated for majority and minority spin channels are $\langle V_{F\uparrow}\rangle = 4.3\text{x}10^7$ cm/s, $\langle V_{F\downarrow}\rangle = 1.1\text{x}10^7$ cm/s respectively.

The self-consistency of our estimates of the mean free path $L$ from the temperature dependence of the contact resistance above ($L$= 330 nm) can be checked against our theoretical results by calculating the conductance, $\sigma_i = e^2\left\langle N(E_F)V_{Fi}^2\right\rangle\tau$, where the relaxation time $\tau = L/<V_F>\sim 8\cdot 10^{-13}$ s. Using the calculated values for $\left\langle N(E_F)V_{F\uparrow}^2\right\rangle = 2.2\,\text{Ry}\cdot\text{Bohr}^2$ and $\left\langle N(E_F)V_{F\downarrow}^2\right\rangle = 0.3\,\text{Ry}\cdot\text{Bohr}^2$ in plane we obtain $\left\langle N(E_F)V_F^2\right\rangle = \left\langle N(E_F)V_{F\uparrow}^2\right\rangle + \left\langle N(E_F)V_{F\downarrow}^2\right\rangle = 2.5$ Ry·Bohr$^2$, yielding $\rho \sim 0.25$ μΩ·cm for the low temperature resistivity, which is fairly close to the experimentally measured value of $\rho \sim 0.4$ μΩ·cm.

In summary, we have measured the spin polarization of both type A and type B MnAs epitaxial films grown on (001) GaAs substrate using the PCAR technique. We conclude from temperature-dependent studies of the contact resistance and the resistivity of MnAs that all the point contact characteristics were taken in the purely ballistic regime. This is further supported by the phonon spectrum of Pb, acquired simultaneously with the conductance measurements. The transport spin polarization was measured in the hexagonal plane for type A samples and was determined to be 49% for type A and 44% for type B samples. The somewhat lower spin polarization value measured for type B samples suggests the presence of the spin anisotropy. The experimental results ($P_{1\perp}$ = 49%) are in good agreement with our density functional calculations, with $P_{1\perp}$ = 42% in the hexagonal MnAs plane. In addition, our calculations in the Ohmic (diffusive) regime indicate that 80% of the current in the hexagonal plane of MnAs is transferred by majority spins, further emphasizing the potential importance of this attractive material system for spintronics applications.

This work was supported by DARPA through ONR grant N00014-02-1-0886 and NSF career grant (B.N.) and ONR N00014-99-1-0071, –0716, and N00014-99-1-1093 (N.S.). We thank D.N. Basov, A.G. Petukhov, and I. Zutic for discussions and L.E. Wenger for the use of his magnetic measurement facilities.

Figure captions:

Fig. 1. Normalized conductance of a series of Sn point contacts with different contact resistance $R_c$ (from 29Ω to 44Ω) of a type A MnAs with at $T$ = 1.2 K. For convenience, a uniform vertical offset is used for plotting each consecutive curve.

Fig.2. Normalized conductance data fitted (solid curve) with the model of Ref. 24 at $T$ =1.2 K. (a) Sn point contact with type A MnAs with contact resistance $R_c$ = 37Ω from Fig.1. Fitting parameters: $Z$= 0.1, $P$= 51%); (b) Normalized conductance curve of a Pb point contact with type B MnAs and the corresponding fit with $Z$ = 0.15, $P$ = 52%.

Fig.3. Sn-MnAs point contact resistance of and in-plane resistivity data for the same sample of type A MnAs in the temperature range (~200-240K) where both dependencies are approximately linear. Inset: Resistivity of MnAs between 4K and 300K.

Fig. 4. Phonon spectra for Pb/MnAs (type A) junction. The phonon peaks of lead are shifted to higher energies by the value of the superconducting gap of Pb. Upon subtracting the value of the gap, the two peak energies (symmetric with respect to zero bias) for the phonon spectra of Pb ($V$= 4.65 mV and $V$=8.8 mV) practically coincide with the tabulated peak values of the of the Eliashberg function in Ref. 34.

Fig. 5. Calculated densities of states (solid line, in st/Ry) and in-plane and c-axis electron velocities (dashed lines, in $10^6$ cm/s) for spin-up and spin-down bands of MnAs (top and bottom panels respectively).

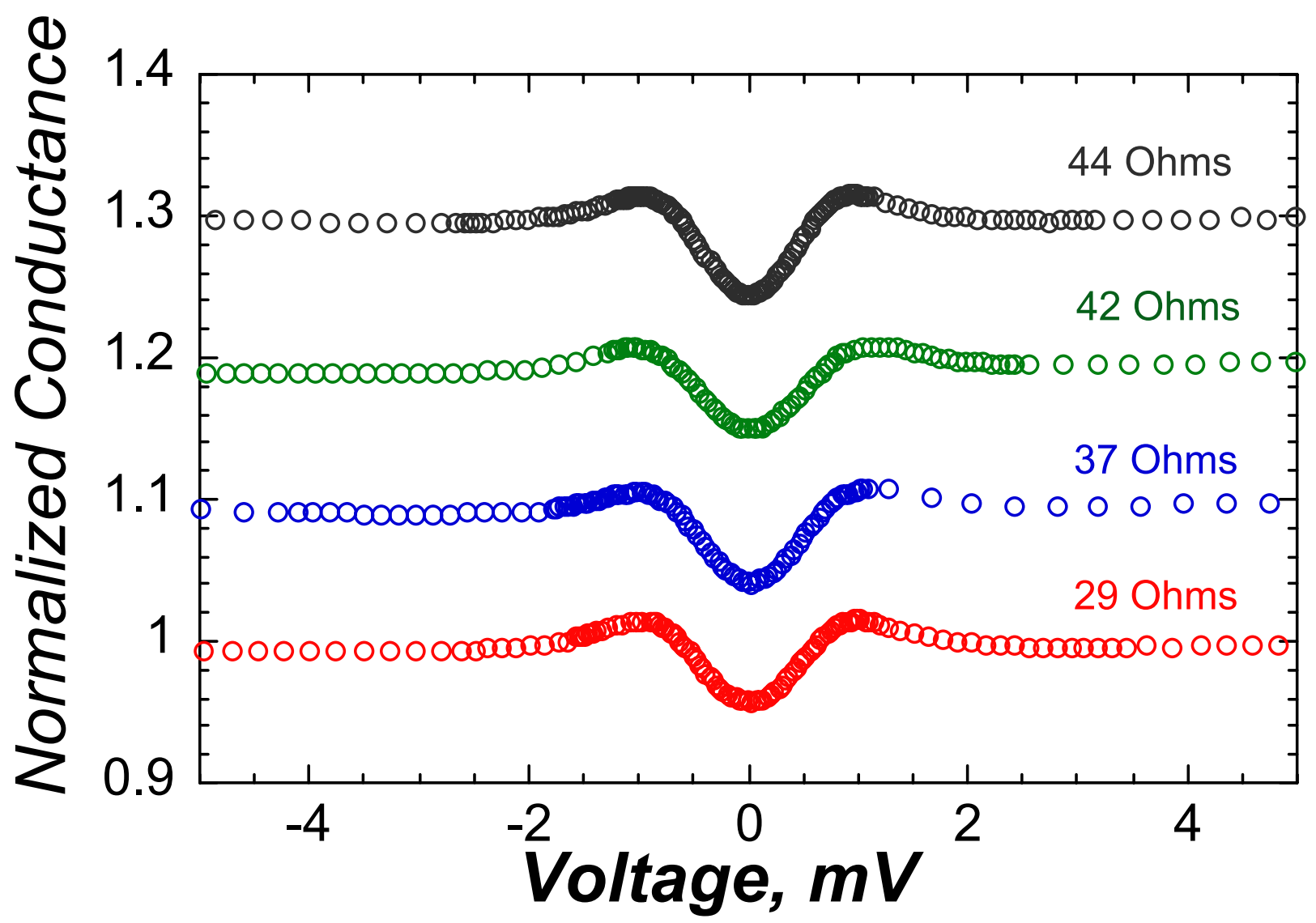

**Fig. 1** **Panguluri et al.**

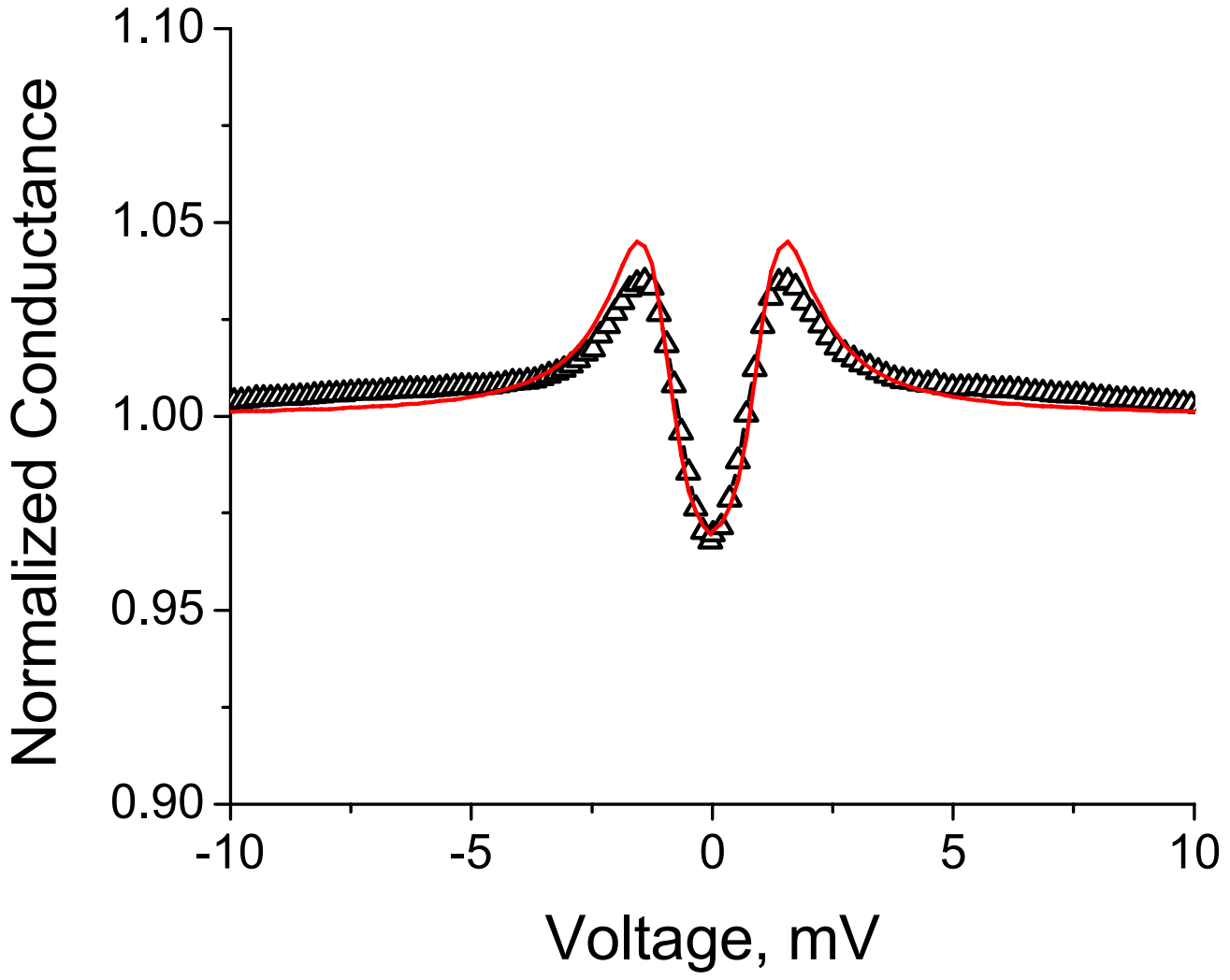

**(a)**

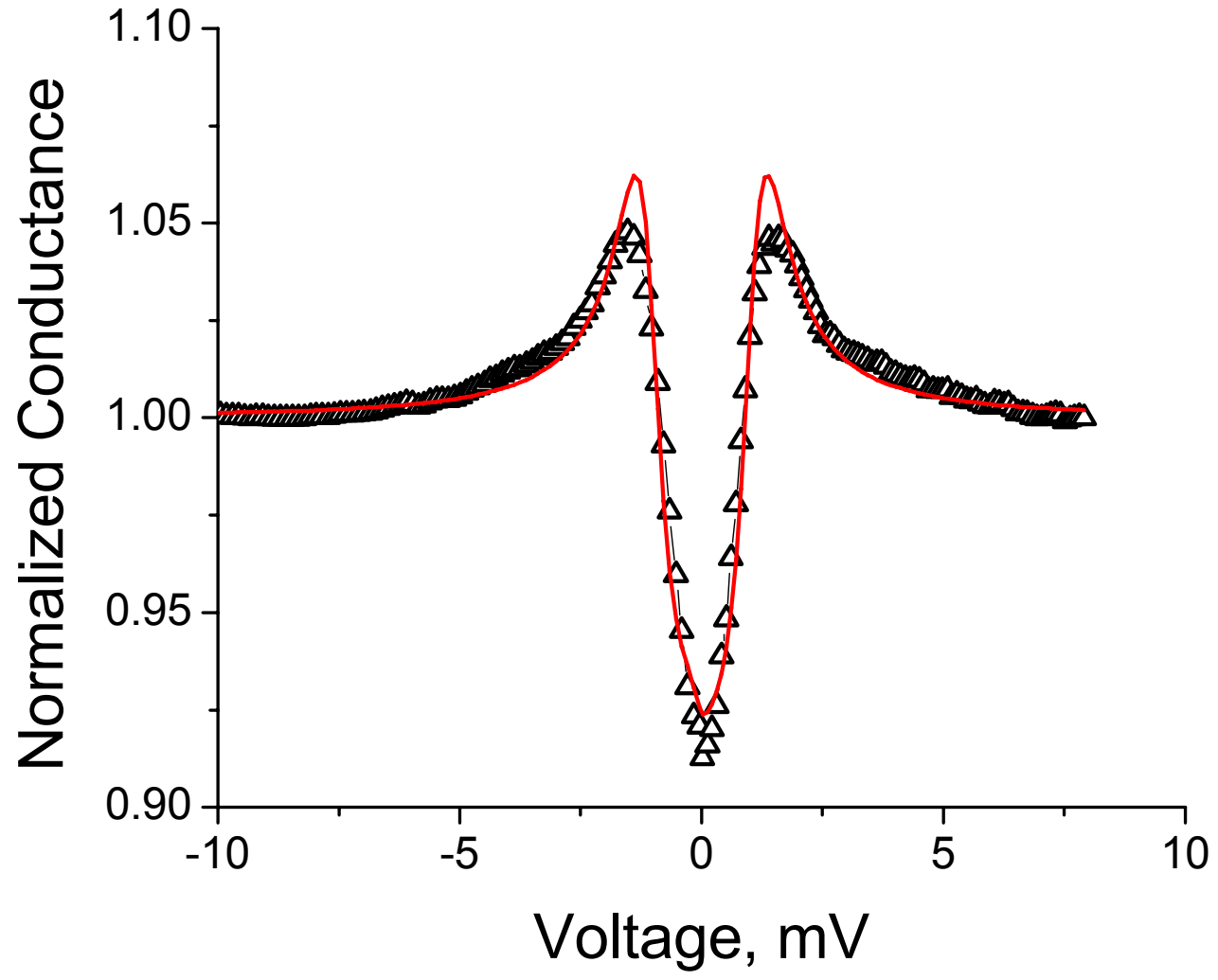

**(b)**

**Fig. 2** **Panguluri et al**

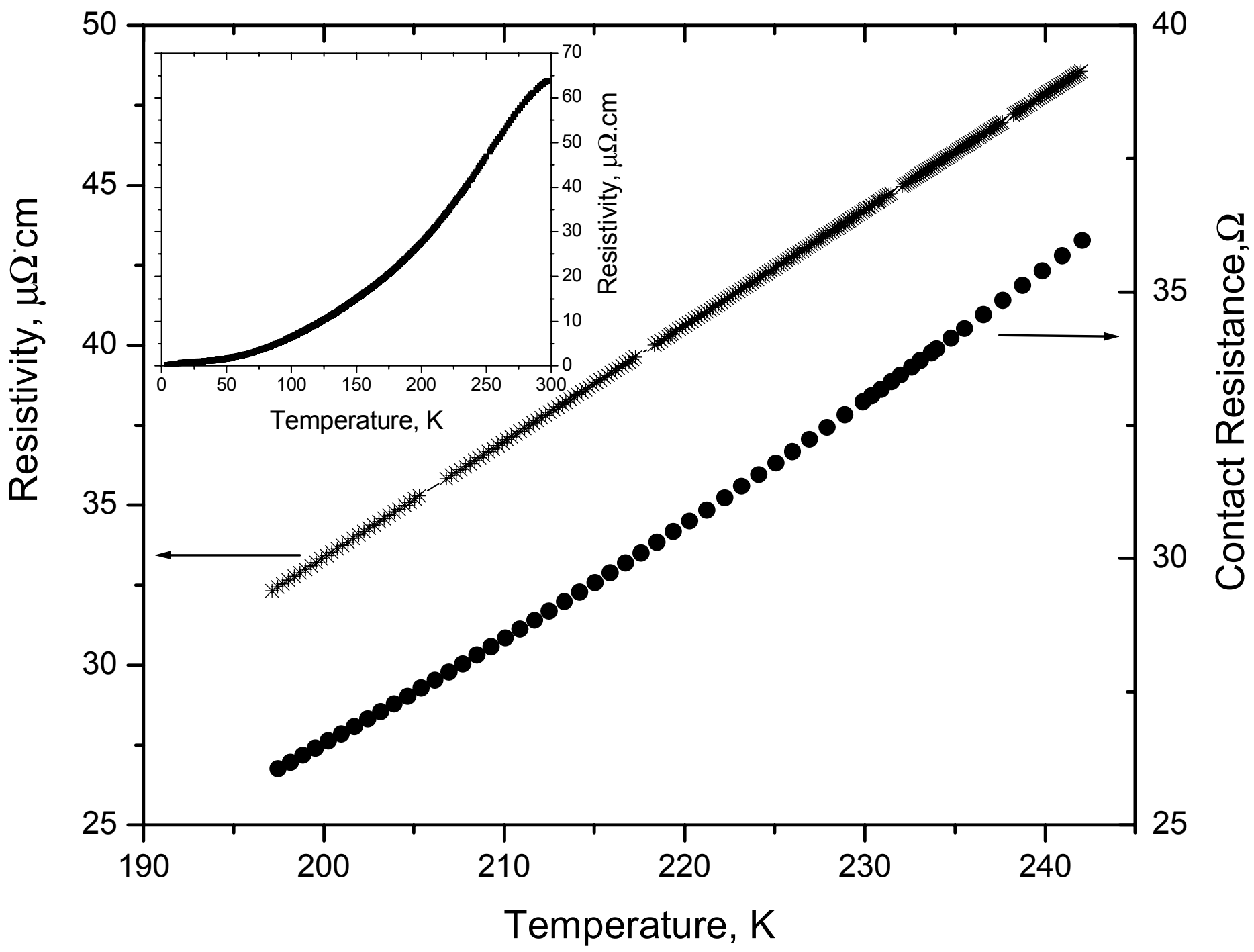

Fig. 3 **Panguluri et al**

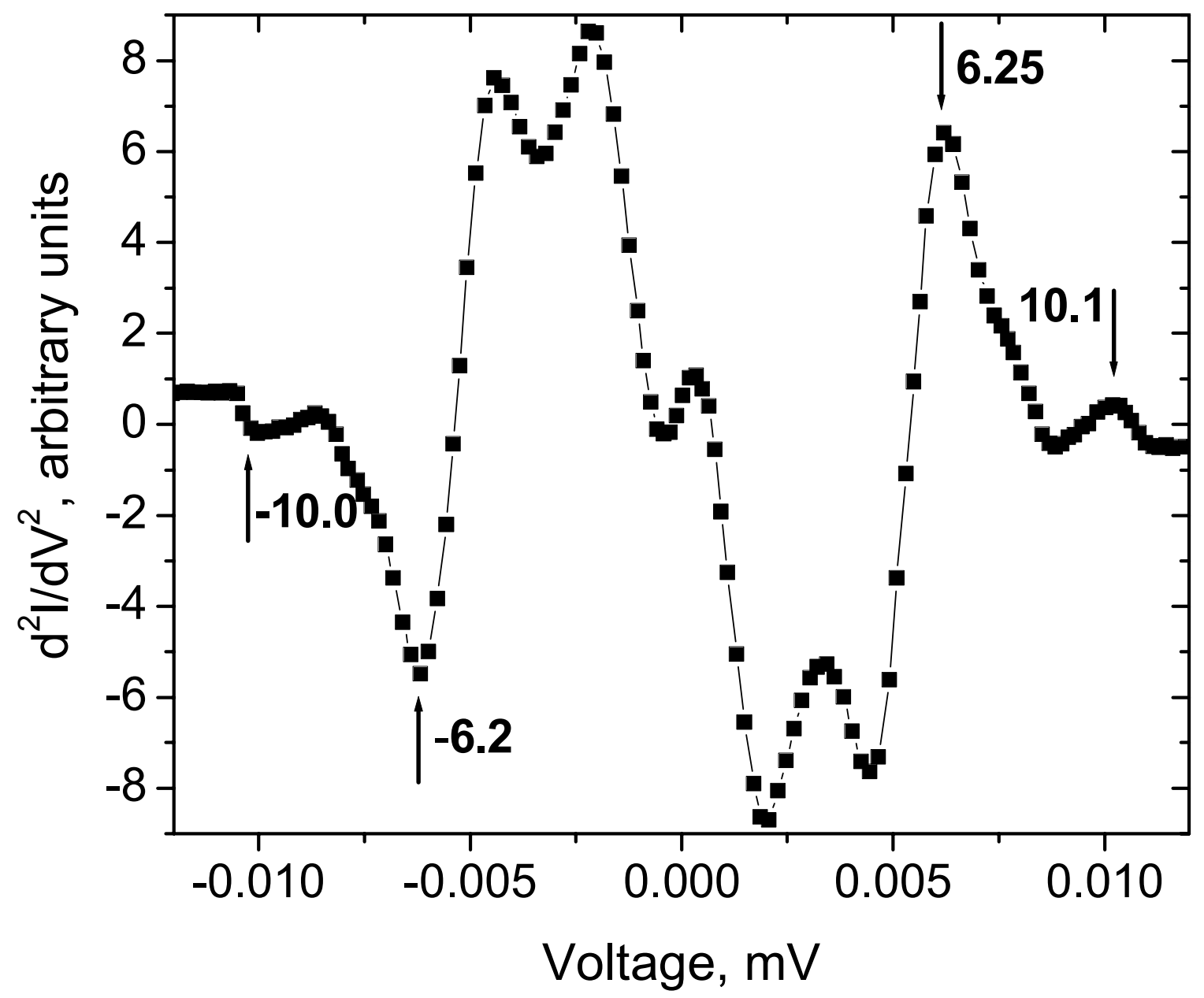

**Fig. 4** **Panguluri et al.**

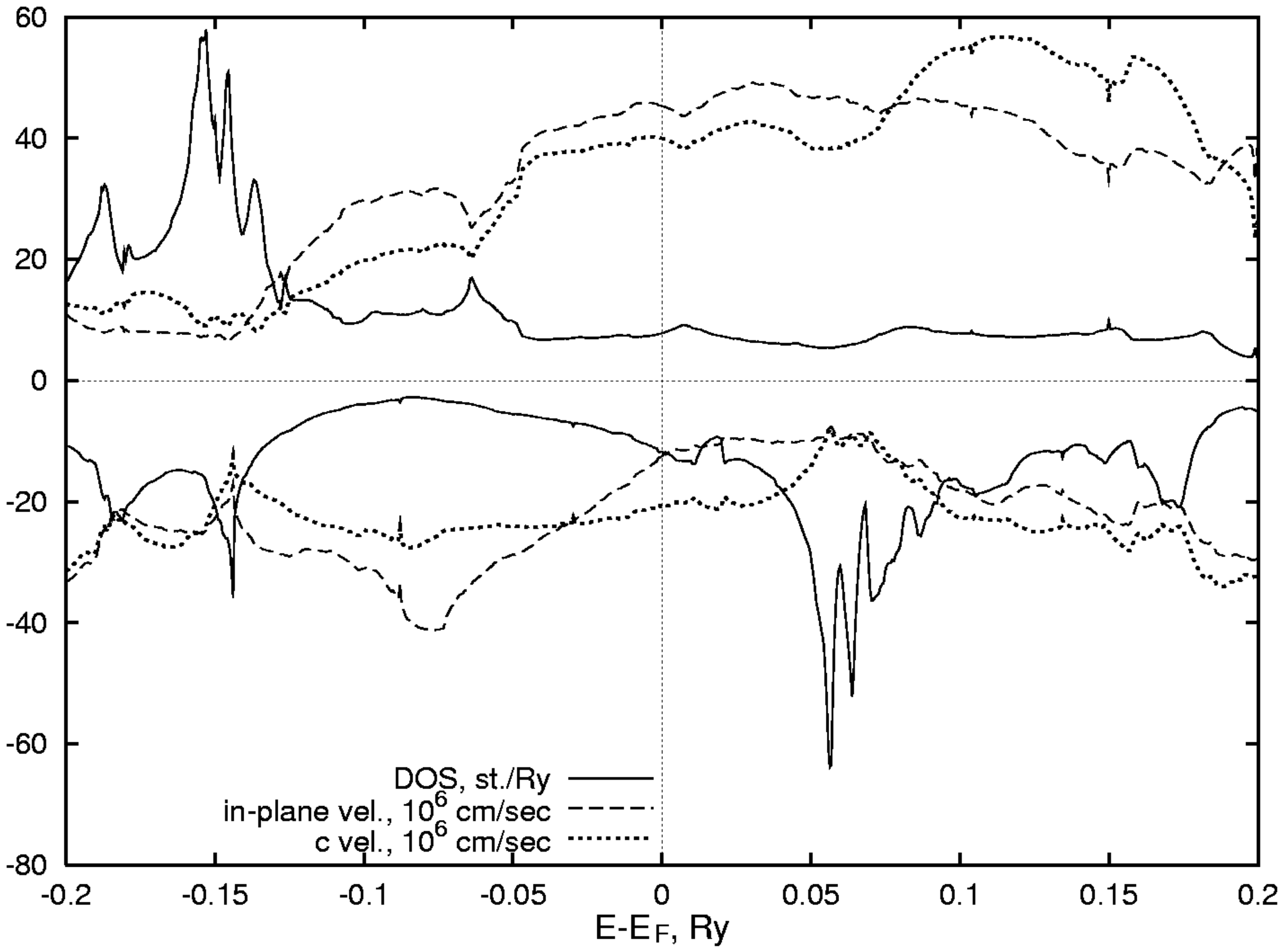

**Fig. 5** **Panguluri et al.**

REFERENCES:

[1]. G.A. Prinz, Physics Today **48** (4), 58 (1995); G.A. Prinz, *Science* **282**, 1660 (1998).

[2]. *Semiconductor Spintronics and Quantum Computation,* D. D. Awschalom, D. Loss, and N. Samarth, Eds. , Springer-Verlag, 2002.

[3]. S.A. Wolf, D.D. Awschalom, R.A. Buhrman, J.M. Daughton, S. von Molnar, M.L. Roukes, A.Y. Chtchelkanova and D.M. Treger*, Science* **294**, 1488 (2001).

[4]. A.G. Aronov, Pis'ma Zh. Expr. and Theor. Fiz **24**, 37 (1976) [*Sov. Phys. JETP Lett.* **24**, 32 (1976)].

[5]. *Optical Orientation*, F. Meier and B.P. Zacharchenia, Eds., North Holland, Amsterdam, 1984.

[6]. D. D. Awschalom and J. M. Kikkawa, Physics Today **52**, 33 (1999) and references therein.

[7]. R. Fiederling, M. Keim, G. Reuscher, W. Ossau, G. Schmidt, A. Waag, L.W. Molenkamp, *Nature* (London*)* **402,** 787 (1999); Y. Ohno, D.K. Young, B. Beschoten, F. Matsukura, H. Ohno, D.D. Awschalom, *ibid* **402,** 790 (1999); B. T. Jonker, Y.D. Park, B.R. Bennett, H.D. Cheong, G. Kioseoglou, A. Petrou, *Phys. Rev. B* **62**, 8180 (2000).

[8]. I. Malajovich, J.J. Berry, N. Samarth, and D.D. Awshalom, Nature (London) **411**, 770 (2001).

[9]. M. Johnson and R.H. Silsbee, Phys. Rev. B **35**, 4559 (1987).

[10]. A. G. Aronov and G.E. Pikus, Fiz. Tech. Poluprovodn. **10**, 1177 (1976). [Sov. Phys. Semicond. **10**, 698 (1976).

[11] P.R. Hammar, B.R. Bennett, M.J. Yang, M. Johnson, Phys. Rev. Lett. **83**, 203 (1999); F.G. Monzon, H.X. Tang, M.L. Roukes, Phys. Rev. Lett. **84**, 5022 (2000); B.J. van Wees, *ibid*, p. 5023; P.R. Hammar, B.R. Bennet, M.J. Yang, M. Johnson, *ibid,* p. 5024.

[12] P.C. van Son, H. van Kempen, and P. Wyder, *Phys. Rev. Lett.* **58** 2271 (1987); G. Schmidt *et al., Phys. Rev*. B **62**, R4790 (2000).

[13]. S. H. Chun, S. J. Potashnik, K.C. Ku, P. Schiffer, and N. Samarth, *Phys. Rev*. B **66**, 100408 (2002).

[14]. A.T. Hanbicki, B.T. Jonker, G. Itskos, G. Kioseoglou, A. Petrou., *Appl. Phys. Lett.*

**80**, 1240 (2002); A. T. Hanbicki *et al*., Appl. *Phys. Lett*. **82**, 4092 (2003).

[15]. E.I. Rashba, *Phys. Rev*. B **62** R16267 (2000); D. L. Smith and R. N. Silver, *Phys. Rev*. B **64**, 045323 (2001).

[16]. Y.-J Zhao, W.T. Geng, A.J. Freeman, and B. Delley, *Phys. Rev*. B **65**, 113202 (2002). According to Ref. 16 the zinc-blend MnAs is only predicted to be a half-metal at a greater than equilibrium lattice structure (5.8 Å instead of 5.7 Å).

[17]. S. Sanvito and N. A. Hill, *Phys. Rev*. B. **62**, 15553 (2000).

[18]. M.Tanaka, J.P. Harbison, M. C. Park, Y. S. Park, T. Shin, and G. M. Rothberg, *Appl. Phys. Lett.* **65**, 1964 (1994).

[19]. J. J. Berry, S. J. Potashnik, S.H. Chun, K.C. Ku, P. Schiffer, and N. Samarth, *Phys*, Phys. Rev. B **64**, 052408 (2000).

[20]. R.J. Soulen, Jr., J.M. Byers, M.S. Osofsky, B. Nadgorny, T. Ambrose, S.F. Cheng, P.R. Broussard, C.T. Tanaka, J. Nowak, J. S. Moodera, A. Barry, and J.M.D. Coey, *Science*, **282**, 85-88 (1998).

[21]. S.K. Upadhyay, A. Palanisami, R.N. Louie, and R.A. Buhrman, *Phys. Rev. Lett*. **81**, 3247-3250 (1998).

[22]. A. F. Andreev, *Sov. Phys. JETP* **19**, 1228 (1964).

[23]. M.J.M. de Jong and C.W.J. Beenakker, *Phys. Rev. Lett*. **74**, 1657-1660 (1995).

[24]. I.I. Mazin, AA. Golubov and B. Nadgorny*, Journ. Appl. Phys.* **89**, 7576 (2001).

[25]. Blonder, G. E. Tinkham, and T.M. Klapwijk, *Phys. Rev.* **B 25**, 4515 (1982).

[26]. K.Xia, P.J. Kelly, G.E.W. Bauer, and I Turek, *Phys. Rev. Lett.,* 89, 166603 (2002).

[27]. I. Zutic and O.T. Valls, *Phys. Rev*. B **61** 1555 (2000).

[28]. B. Nadgorny, I.I. Mazin, M. S. Osofsky, R. J. Soulen, Jr., P. Broussard, R.M. Stroud, D.J. Singh, V.G. Harris, A. Arsenov, and Ya. Mukovskii, *Phys. Rev.* ***B*****. 63,** 184433 (2001).

[29]. F. Schippan, G. Behme, L. Daweritz, K. H. Ploog, B. Dennis, K.-U. Neumann, and K. R. A. Ziebeck, *J. Appl. Phys*. **88**, 2766 (2000).

[30]. Y. Ji Y, G.J. Strijkers, F.Y. Yang, C.L. Chien, J.M. Byers, A. Anguelouch, G. Xiao, A. Gupta, *Phys. Rev. Lett.* **86**, 5585 (2001); C.H. Kant, O. Kurnosikov, A.T. Filip, P. LeClair, H.J.M. Swagten, and W.J.M. de Jonge, *Phys. Rev*. B **66,** 212403 (2002).

[31]. The spin polarization in both the ballistic and the diffusive regimes is higher in the hexagonal plane than perpendicular to it (along the c-axis of the MnAs). For type A MnAs the current is measured in-plane, whereas for the type B the current is measured at an angle of approximately 23° to the hexagonal plane.

[32]. A.I Akimenko, A.B. Verkin, N.M. Ponomarenko, and I.K. Yanson, *Sov. Journ. Low Temp. Phys*. **8**, 130 (1982).

[33]. In Ref. 32, the base and the probe materials were the same. Here we assume that the major contribution to the contact resistance comes from MnAs, which is confirmed by our measurement of the resistivity of MnAs and Sn in the appropriate temperature range. For this experiment Sn is preferable to Pb, as it has a higher Debye temperature and a lower resistivity.

[34]. See A.V. Khotkevich and I.K. Yanson, *Atlas of Point Contact Spectra of Electron-Phonon Interactions in Metals*, Kluwer, Boston, 1995 and the references therein.

[35]. A.A. Lysykh, I.K. Yanson, O.I. Shklyarevskii, and Yu.G. Naidyuk, *Solid State Commun*. **35**, 987 (1980).

[36]. J. P. Perdew and Y. Wang, *Phys. Rev. B* 45, 13244 (1992).

[37]. K. Katoh, A. Yanase, and K. Motizuki, *J. Magn. Magn. Mater*. **54**, 959 (1986); P. M. Oppeneer, V. N. Antonov, T. Kraft, H. Eschrig, A. N. Yaresko, and A. Ya. Perlov, *J. Appl. Phys*. **80**, 1099 (1996); P. Ravindran, A. Delin, P. James, and B. Johansson, J. M. Wills, R. Ahuja and O. Eriksson, *Phys. Rev.* B **59**, 15680 (1999).

[38]. B. Nadgorny, R. Soulen, M.S. Osofsky, I.I. Mazin, G. Laprade, R.J.M. van de Veerdonk, A.A. Smits, S.F. Cheng, E.F. Skelton, S.B. Qadri, *Phys. Rev.* B*,* **61**, R3788 (2000); note that the sign of *P* cannot be experimentally confirmed by the PCAR technique.